\documentclass[12pt, a4paper]{article}
\pdfoutput=1
\usepackage{color}
\usepackage{graphicx}
\usepackage{caption}
\usepackage{amsmath}
\usepackage{subcaption}
\usepackage{mathtools}
\usepackage{url}
\usepackage[font=scriptsize,labelfont=bf,skip=6pt]{caption}
\usepackage[T1]{fontenc}
\usepackage{authblk}

\begin{document}
\title{A new technique to establish the uniformity of the induction gap in GEM based detector}

\author{Mohit Gola\thanks{Corresponding Author: mohit.gola@cern.ch}}
\author[b]{F. Brunbauer}
\author[b]{D. Janssens}
\author[a]{A.~Kumar}
\author[b]{M. Lisowska}
\author[b,c]{H. Muller}
\author[a]{~Md. Naimuddin}
\author[b]{E. Oliveri}
\author[b,d]{D. Pfeiffer}
\author[b]{L. Ropelewski}
\author[b,c]{L. Scharenberg}
\author[b]{M. Van Stenis}
\author[b]{A. Utrobicic}
\author[b]{R. Veenhof}

\newcommand{\dd}[1]{\mathrm{d}#1}

\affil[a]{Department of Physics $\&$ Astrophysics\\ University of Delhi\\ Delhi, India}
\affil[b]{European Organization for Nuclear Research (CERN)\\ CH-1211 Geneve 23\\ Switzerland}
\affil[c]{Physikalisches Institut, University of Bonn\\ Nußallee 12, 53115 Bonn\\ Germany}
\affil[d]{European Spallation Source (ESS ERIC)\\ P.O. Box 176, SE-22100 Lund\\ Sweden}

\maketitle
\begin{abstract}

This work is proposing and exploring the use of multichannel readout electronics, already used in quality assurance for gain uniformity studies, to measure the uniformity of the induction gap in GEM based detectors. The measurement will furthermore provide a qualification of the readout electrodes in terms of disconnected or shorted channels. The proposed method relies on the dependence on the induction gap capacitance between the readout strips and the bottom of the GEM. The measurement is obtained inducing capacitively a signal on the readout strips pulsing the bottom layer of the GEM foil. In this work, the signals are read with the analog APV25 front-end chip and the RD51 Scalable Readout System (SRS). Studies on small and large area GEM detectors, relative variations under mechanical stress, and in presence of electrical fields will be shown. Sensitivity to defects in the readout plane will be proven.
\end{abstract}
\begin{keywords}
MPGD, APV25, GEM, SRS
\end{keywords}

\section{Introduction}

Many existing experiments and their upgrades in High Energy Physics (HEP) are proposing and utilizing Micro Pattern Gas Detectors (MPGDs) to have large area coverage~\cite{0,1,2,2.1}. Quality controls (QA/QCs) are important to validate detectors and achieved performances before they will be installed in experiments~\cite{3,4,5}.  QA/QCs procedure have to be reliable, reproducible and with possibly as many information as needed. Along this line, a method has been developed to qualify the induction gap in a Gas Electron Multiplier (GEM) based detector~\cite{6,7,8,9}. The flatness of the induction gap is an important parameter because the gain of the GEM detector is linearly dependent upon it. Non-uniformity in this gap causes variation of gain over the surface (i.e. the gain uniformity) affecting overall energy resolution, dynamic range of the signal, compatibility with the electronics, and detector stability. This work proposes the use of multichannel analog readout chips, already in use for gain uniformity qualification, to measure the uniformity of the induction gap. Direct comparison of these two measurements will be very beneficial and it will not add complication to existing QA/QCs procedure. The result of this measurement does not only help in QA/QCs procedure but also helps in R\&D measurements in the validation of the prototypes.\\
The structure of the paper is as follows: section \ref{Methodology} explains the technique developed and described in this paper, section \ref{Setup} explains the experimental setup. Section \ref{Measurements} describes the set of measurements performed in order to validate and demonstrate the capability of this method. Finally, we will conclude our findings.

\section{Methodology}
\label{Methodology}

The method used in this paper relies on the indirect measurement of the gap through the capacitance between readout electronics and bottom GEM.  An external pulse is sent to the bottom of the GEM, and the amplitude of the capacitively induced signal on readout strips is measured. The measurement can be done with or without high voltage applied to the detector. This method requires a multichannel analog front-end chip and it should be calibrated to disentangle the non-uniformity of the chip with the detector. The caveats associated with this technique that it is sensitive to all the types of coupling. In the paper we will show for instance the effect of the readout strips fan-out. The method, being sensitive to the electrode capacitance, can be used to qualify readout electrodes. Shorts or missing connections will be identified for instance because of the anomalous electrode capacitance.

\section{Setup}
\label{Setup}

\begin{figure}[h]
\centering
\includegraphics[width=0.5\textwidth]{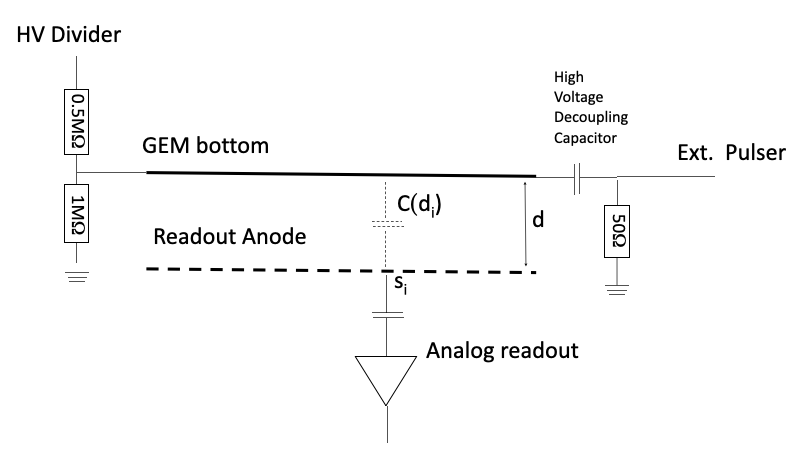}
\caption{Shows the schema for the technique describe in this paper. }
\label{fig:Schema}
\end{figure}

The setup shown in the Figure~\ref{fig:Schema}, can divided into two parts: the charge injection circuity and the multichannel readout system based on APV25 and SRS~\cite{10,11}. The external pulse has been generated using the 25 GHz pulse generator with a definite pulse frequency of 1 kHz, amplitude 1.15 V peak to peak, width 1 $\mu$S, and delay of 300 ns. The signal from the pulse generator fed into the bottom of the GEM foil in series with 100 pF capacitor and a 50 $\Omega$ termination. The external pulser has been synchronized with the SRS clock using a synchronous NIM signal produced by the SRS.

\begin{figure}[h]
\centering
\includegraphics[width=0.5\textwidth]{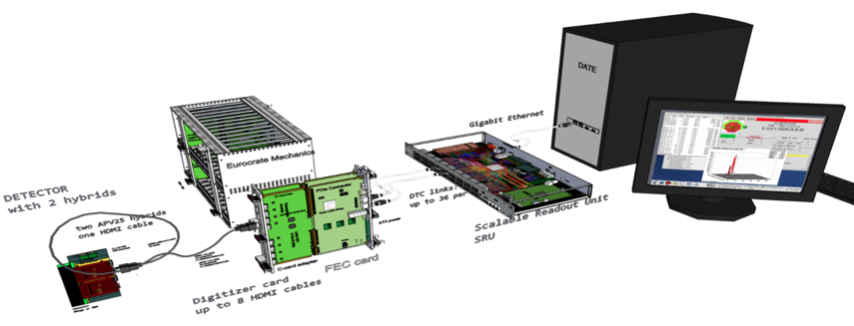}
\caption{Shows the schematic os the Scalable Readout System(SRS). }
\label{fig:SRS}
\end{figure}

The data was recorded using the SRS shown in Figure~\ref{fig:SRS}, which has been developed by the RD51 collaboration at CERN. The CMS GEM collaboration use this system for the quality assurance of the GEM detector~\cite{13,14,15}. 

\begin{figure}[h]
    \begin{subfigure}[b]{0.4\textwidth}
    \centering
    \includegraphics[width=\textwidth]{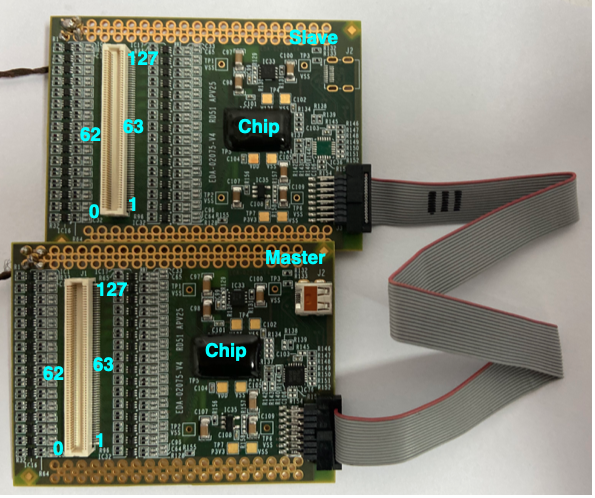}
    \caption{}
    \end{subfigure}
    \begin{subfigure}[b]{0.6\textwidth}
    \centering 
    \includegraphics[width=\textwidth]{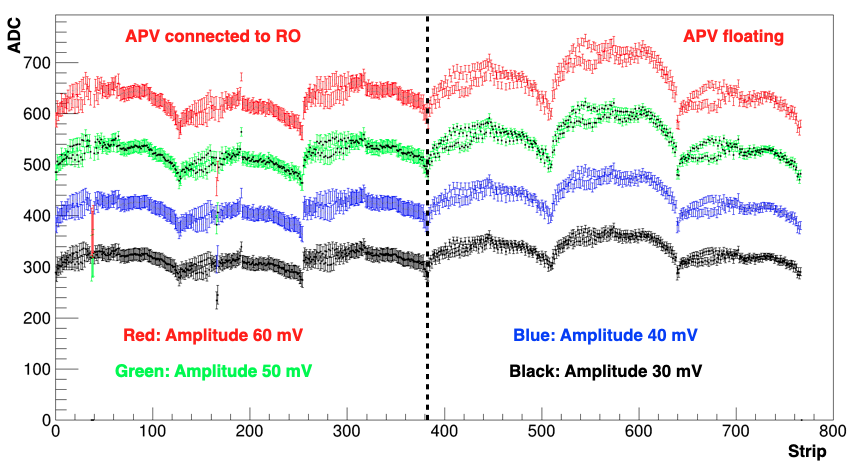}
    \caption{}
    \end{subfigure}
    \caption{ (a) Shows the pair of APV chip , and (b) Shows the behaviour of APV-pair operated in calibration mode for different amplitude input pulse.} 
\label{fig:ASIC}
\end{figure}

The SRS has been used to control and read the APV25 front end ASIC. APV25 has 128 channels that will be digitized with a sampling clock of 40MHz as shown in the Figure~\ref{fig:ASIC} (a), the details of the chip can be found elsewhere~\cite{16,17}. In order to evaluate non uniformities intrinsic to the chip and the RD51 hybrids, a calibration scan with internal pulses has been done. For the calibration, 16 strips has been fired at the same time and data have been later combined. Figure~\ref{fig:ASIC} (b) shows the ADC value recorded for the different amplitudes of the input pulse. For this, three pairs of master and slave APV has been used, out of which master APVs were connected to the readout board and slave APVs were left floating. A similar pattern has been observed for both the APVs either connected to readout (RO) or left floating hence there is finite non-uniformity present in the channels of the chip.

\section{Measurements}
\label{Measurements}

A series of measurements have been carried out on the small and large size GEM detectors in order to extract the information regarding the flatness of the induction gap. Also, the application of this technique has been described in terms of deformation of the induction gap, quality assurance of the readout board, etc.

\subsection{Sensitivity of the induction gap}

The aim of this measurement is to check the sensitivity of the signal gap utilizing the capacitance of the gap. The experimental setup for this measurement uses the two-dimensional RO board having an active area of 10 cm X 10 cm and consists of 256 strips in X-direction and the same in Y-direction.

\begin{figure}[h]
        \hspace{1 cm}
    \begin{subfigure}[b]{0.25\textwidth}
    \centering
    \vspace{0.2 cm}
    \includegraphics[width=\textwidth]{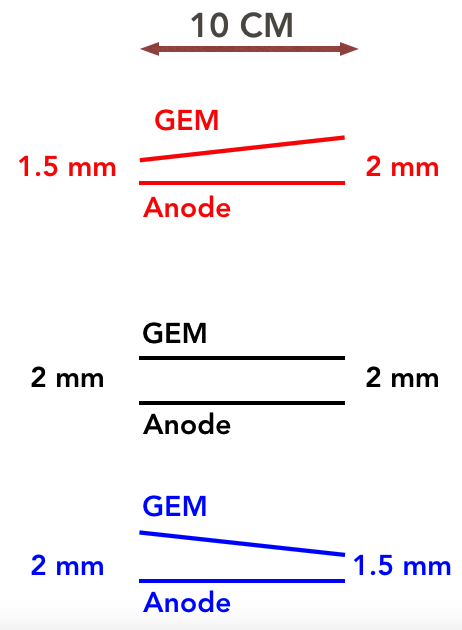}
    \caption{}
    \end{subfigure}
    \hspace{1 cm}
    \begin{subfigure}[b]{0.575\textwidth}
    \centering 
    \includegraphics[width=\textwidth]{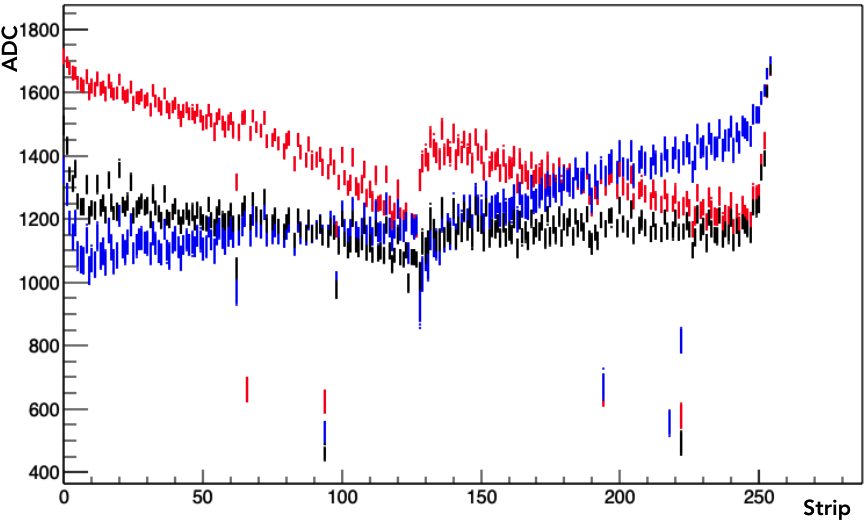}
    \caption{}
    \end{subfigure}
    \caption{ (a) Shows the different orientations for GEM \& Anode plane, and (b) shows the ADC value for each strip for all the three configurations.} 
\label{fig:POP}
\end{figure}

A GEM foil having an active area similar to the RO and it's placed on top of the RO board~\cite{18,19}. The output of the pulse generator has been connected to the bottom of the HV trace of GEM foil. Panasonic connectors in X-direction were equipped with the APV25 chip and Y-direction strips were terminated with 0 $\Omega$. Three different gap configurations were tested in order to look for the induction gap sensitivity as shows in the Figure~\ref{fig:POP} (a). Firstly, a gap of 2 mm was maintained between the bottom of the GEM foil and anode then the foil was titled with the asymmetric gap of 1.5 mm on the left and 2 mm on the right, and vice-versa. The Figure~\ref{fig:POP} (b) shows a clear variation in the ADC value with respect to strip has been observed for the three different gap configurations.

\subsection{Large area triple-GEM detector}

After testing this technique on 10 cm $\times$ 10 cm GEM detector, it has been applied to the large ($O (m^2)$) size triple-GEM detector with known non-uniformity in the induction gap. For this measurement, a triple-GEM detector having a physical dimension of 1 m $\times$ 55 cm $\times$ 24 cm with a gap configuration of 3 mm/1 mm/2 mm/1 mm for the Drift/Transfer1/Transfer2/induction gap, respectively has been used. The readout board for this detector is 3 mm thick PCB having in total 3072 one-dimensional strips and having known deformation in the detector. 

\begin{figure}[h]
\centering
\includegraphics[width=0.7\textwidth]{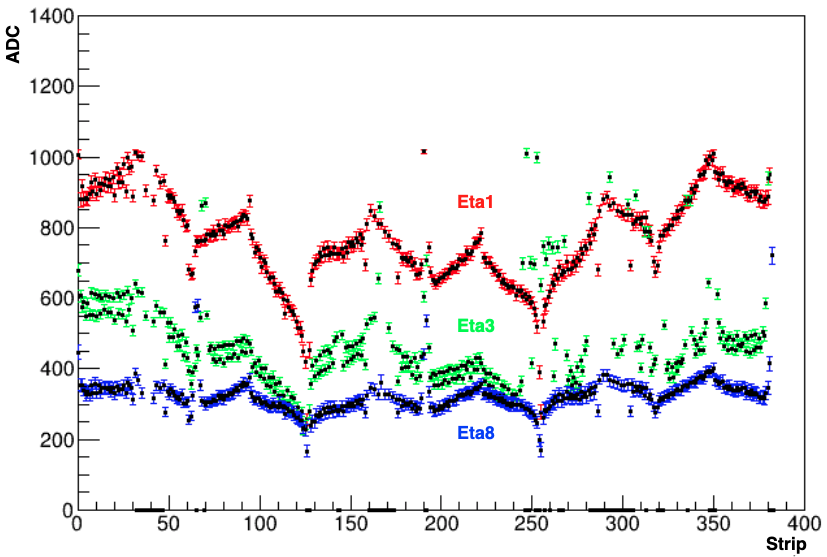}
\caption{Shows the ADC value for each strip and 3-Eta segments of the triple-GEM detector. }
\label{fig:GE11}
\end{figure}

\begin{figure}[h]
\centering
\includegraphics[width=0.8\textwidth]{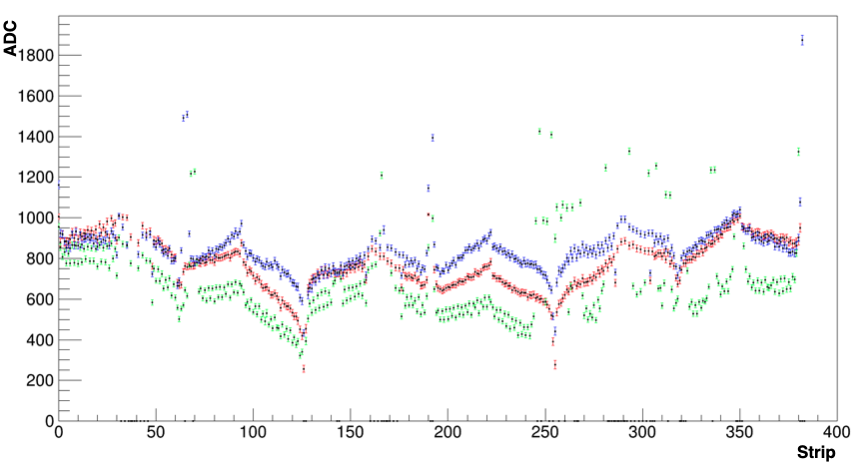}
\caption{Shows the capacitance correction applied on the three particular Eta sectors namely Eta1, Eta3 and Eta8. }
\label{fig:GE11_Cap}
\end{figure}

The RO board is divided into 8 eta segment and 3 phi partitions such that each (eta, phi) sector has 128 strips. RO board has different strip lengths for each eta segment as going from the wider side to the narrow side. A high voltage divider has been used to power up the GEM foils and gaps simultaneously~\cite{20}. An external pulse was sent to the bottom of the third GEM foil in parallel to the divider, and a scan of three eta segment has been performed as shown in the Figure~\ref{fig:GE11}. It shows that the detector is not completely flat and having a clear manifestation of the bedding (outward) in the RO board and shows the fan-out effect of the readout strips.  Due to different strip length and width in each eta sector, a correction on capacitance has been applied for two particular eta sector with respect to the eta sector on the wider side as shown in the Figure~\ref{fig:GE11_Cap}.

\subsection{Deformation of the induction gap}
\begin{figure}[h]
\centering
\includegraphics[width=0.8\textwidth]{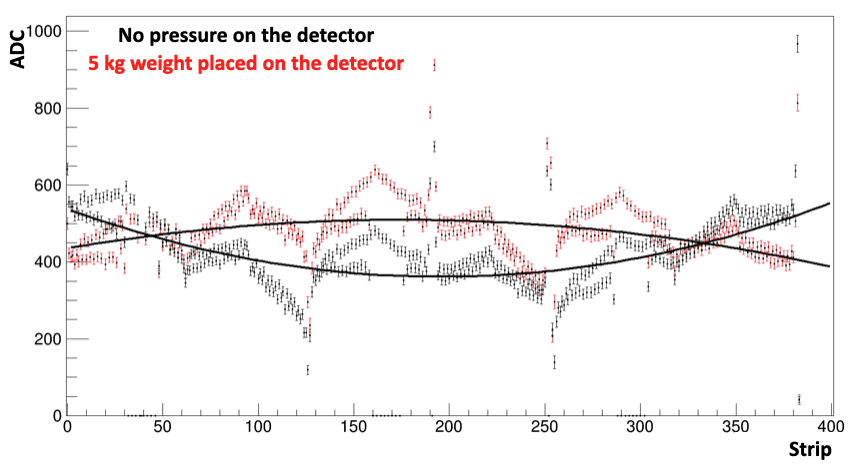}
\caption{Shows the ADC value for each strip for one of the Eta segment With and without 5kg weight placed. }
\label{fig:Weight}
\end{figure}

In the next test, we deformed the gap and the detector on purpose. In order to demonstrate this a weight of roughly 5 kg has been placed on one of the eta segment and look for the ADC value with respect to the strip. The two plots i.e. with weight placed on RO and without are plotted together to see the difference as shown in the Figure~\ref{fig:Weight}. The plot with weight having a higher ADC value for the particular area, where the weight was placed, hence shows the bending of the RO board in the opposite direction i.e. inward. The distribution for both the measurements is fitted with a $\rm 2^{nd}$ order polynomial function to visualize the deformation of the RO board.

\subsection{Test in presence of the induction field} 

The method has been tested as well with high voltage applied, to prove that it can done as well when all the electrical fields are applied. To understand this a 700 V has been applied to the bottom of the third GEM foil and at the same divider pad, an external pulse was also sent. 

\begin{figure}[h]
\centering
\includegraphics[width=0.8\textwidth]{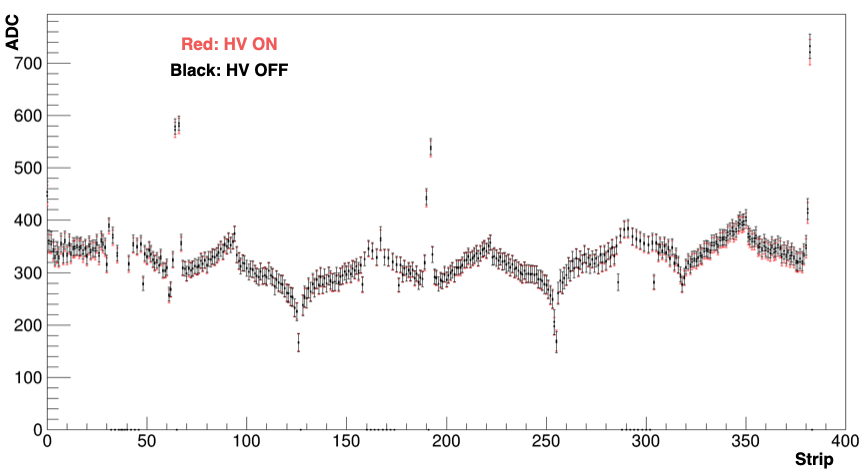}
\caption{Shows the ADC value for each strip for one of the Eta segment when HV is ON and OFF. }
\label{fig:HV}
\end{figure}

Measurement has been carried when the high voltage is ON as well as OFF and two spectra were plotted together as shown in the Figure~\ref{fig:HV}. No variation in the ADC spectrum with respect to strip number observed, hence it shows that this technique works in the presence of high voltage and can be used to monitor the if electrostatic force will induce variation in the gap. Also, this measurement confirms that the mechanical stretching used to stretch the foils is sufficient to overcome the sag due to running high voltage.

\subsection{Readout electrodes integrity} 

\begin{figure}[h]
\centering
\includegraphics[width=0.8\textwidth]{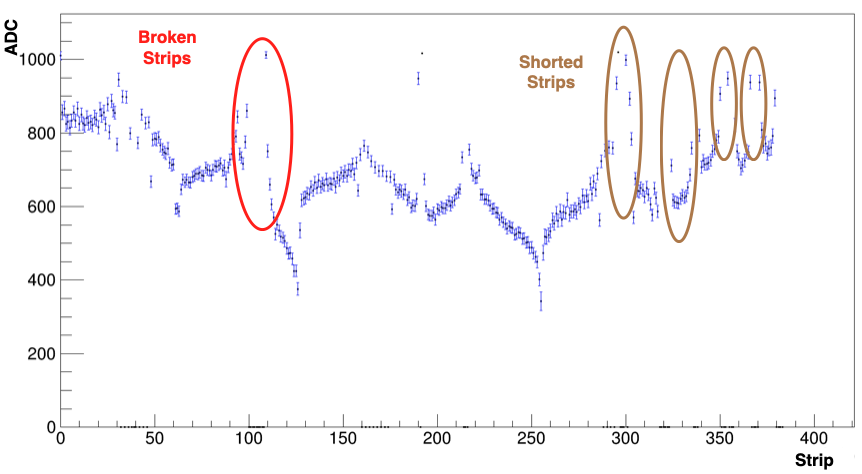}
\caption{Shows the ADC value for each strip for one of the Eta segment having defects present in the strips of the RO. }
\label{fig:Defects}
\end{figure}

The advantage of using a multichannel front-end chip is to locate the defect(s) present in each strip. Measurement has been carried in the readout sector where few strips shorted as well as broken. The ADC spectrum with respect to strip has been plotted as shown in the Figure~\ref{fig:Defects} and observed that due to change in capacitance for shorted, broken, and damaged strips, a higher value of ADC was observed. And compared to the physical location of the bad strips on the RO board. Hence this technique can be used to do the quality assurance of the readout boards well before using them in detector assembly.

\section{Conclusion}

We explored a measurement technique sensitive to the local size of the induction gap of GEM based detectors. In this work the measurement has been performed using the RD51 SRS/APV25 as analog multichannel readout system. The aim of this work was to give a proof of principle of the technique. Potential as been shown as well limiting factors as spurious effects coming for instance from additional capacitances not dependent in the gap as the one associated to the strips fanout. For a proper and quantitative analysis, aspects like this one have to be properly taken into account. An outstanding feature of this technique is that it can be used under the operation of the high voltage which means that this technique can be applied at the field configuration used in the experiment. Also, it can be used for the quality assurance of the readout boards manufactured by the industries to detect possible defects during production. This technique can be used for any gaseous detector properly couple a plane to any readout.
 
\section{Acknowledgements}
We would like to acknowledge the funding agency, Department of Science and Technology (DST), New Delhi (grant nos. SR/MF/PS-02/2014-DUA(G), and SB/FTP/PS-165/2013) for providing financial support. We would also like to thanks the RD51 collaboration at CERN to provide the facility and support.


\begin{thebibliography}{99}

\bibitem{0}
S. D. Pinto et al., "Progress on large area GEMs", JINST 4 P12009 (2009). 

\bibitem{1}
CMS Collaboration, "Overview of large area triple-GEM detectors for the CMS forward muon upgrade", NIMA 845 (2017) 298-303.

\bibitem{2}
P. Karchin, Performance of a Large-Area Triple-GEM Detector in a Particle Beam, Physics Procedia, 37 (2012) 561-566.

\bibitem{2.1}
D. Abbaneo,  \textbf{M. Gola} et al., "Performance of GE1/1 Chambers for the CMS Muon Endcap Upgrade", arXiv:1903.02186.

\bibitem{3}
CMS collaboration, "Quality control and beam test of GEM detectors for future upgrades of the CMS muon high rate region at the LHC", JINST (2015) 10 C03039.

\bibitem{4}
\textbf{M. Gola} on behalf of CMS Collaboration; Muon Chamber End-cap Upgrade of the CMS Experiment with Gas Electron Multiplier (GEM) Detectors and their Performance; Springer Proc. in Phy. 203 (2018) 591.

\bibitem{5}
E. Soldani, "Production and quality control of GEM detectors for the phase 1 upgrade of the CMS experiment", IL NUOVO CIMENTO 41 C (2018) 90.

\bibitem{6}
S.D. Pinto 2010 Micropattern gas detector technologies and applications the work of the RD51 collaboration, IEEE Nucl. Sci. Symp. Conf. Rec. 802.

\bibitem{7}
F. Sauli, GEM: A new concept for electron amplification in gas detectors, Nucl. Instrum. Meth. A, 386, 1997: 531, \url{doi: 10.1016/S0168-9002(96)01172-2}.

\bibitem{8}
F. Sauli, The gas electron multiplier (GEM): Operating principles and applications, Nucl. Instrum. Meth. A, 805, 2016; 2, \url{doi: 10.1016/j.nima.2015.07.060}.

\bibitem{9}
F. Sauli, Gas Electron Multiplier (GEM) Detector: Principles of operation and applications, RD51-NOTE-2012-007.

\bibitem{10}
S Martoiu, H Muller, A Tarazon and J Toledo, "Development of the scalable readout system for micro-pattern gas detectors and other applications", JINST 8, 2013.

\bibitem{11}
Sorin Martoiu, Hans Muller, J. Toledo, "Front-end electronics for the Scalable Readout System of RD51", \url{DOI: 10.1109/NSSMIC.2011.6154414}.


\bibitem{13}
A. Shah, A. Ahmad, \textbf{M. Gola}, R. Sharma, S. Malhotra, A. Kumar, Md. Naimuddin, P. Menon, K. Srinivasan, "Development, Characterization and Qualification of first GEM foils produced in India"; NIM A 892 (2018) 10-17.

 \bibitem{14}
 \textbf{M. Gola}, S. Malhotra, A. Shah, A. Ahmed, A. Kumar, Md. Naimuddin, "Performance of the triple GEM detector built using commercially manufactured GEM foils in India"; NIM A 951 (2019). 

\bibitem{15}
 \textbf{M. Gola}, S. Malhotra, A. Kumar, Md. Naimuddin, "Stability test performed on the triple GEM detector built using commercially manufactured GEM foils in India";2019 JINST 14 p08004.

\bibitem{16}
M. Raymond et al., " The APV25 0.25 m CMOS readout chip for the CMS tracker", IEEE Nucl. Sci. Symp. Conf. Rec. 2 9/113.

\bibitem{17}
Lawrence Jones (RAL), "APV25 user guide", CERN CDS-002725643.

\bibitem{18}
A. Bressan et al.,"Two-dimensional readout of GEM detectors", \url{DOI: 10.1016/S0168-9002(98)01405-3}.

\bibitem{19}
A.Bondar et al., "Performance of the triple-GEM detector with optimized 2-D readout in high intensity hadron beam", arXiv 0110054.

\bibitem{20}
J. A. Merlin, Study of long-term sustained operation of gaseous detectors for the high rate environment in CMS, CERN-THESIS-2016-041.




\end{thebibliography}
\end{document}